\pdfoutput=1

\documentclass[11pt]{article}

\usepackage[]{acl}
\usepackage{times}
\usepackage{latexsym}
\usepackage[T1]{fontenc}
\usepackage[utf8]{inputenc}
\usepackage{microtype}
\usepackage{booktabs}
\usepackage{arydshln}
\usepackage{tabu}
\usepackage[inline]{enumitem}
\usepackage{multirow}
\usepackage{array}
\usepackage{xcolor}
\usepackage{amsmath,amssymb,amsfonts}
\usepackage[makeroom]{cancel}
\usepackage{graphicx}
\usepackage{pgf}
\usepackage{trimclip}
\newcommand{\clip}[1]{\clipbox{1.02cm 0.8cm 0.85cm 0.75cm}{#1}}
\usepackage{calc}
\usepackage{balance}
\usepackage{multicol}
\usepackage{tcolorbox}

\newcommand{\projname}[1]{CoRT\textsubscript{#1}}

\renewcommand{\emph}[1]{\textit{#1}}

\newcommand{\mysubsub}[1]{\subsubsection{#1}}

\newlength{\DepthReference}
\newlength{\HeightReference}
\newlength{\Width}%

\definecolor{MyBlue}{rgb}{0.8,0.8,1}
\definecolor{MyGreen}{rgb}{0.8,1,0.8}
\newcommand{\green}[2][MyGreen]%
{%
    \settowidth{\Width}{#2}%
    \setlength{\fboxsep}{0.5pt}%
    \colorbox{#1}%
    {%
        \raisebox{-0.8\DepthReference}%
        {%
                \parbox[b][\HeightReference+\DepthReference][c]{\Width}{\centering\strut#2}%
        }%
    }%
}

\newcommand{\pink}[2][pink]%
{%
    \settowidth{\Width}{#2}%
    \setlength{\fboxsep}{0.5pt}%
    \colorbox{#1}%
    {%
        \raisebox{-0.8\DepthReference}%
        {%
                \parbox[b][\HeightReference+\DepthReference][c]{\Width}{\centering\strut#2}%
        }%
    }%
}

\newcommand{\blue}[2][MyBlue]%
{%
    \settowidth{\Width}{#2}%
    \setlength{\fboxsep}{0.5pt}%
    \colorbox{#1}%
    {%
        \raisebox{-0.8\DepthReference}%
        {%
                \parbox[b][\HeightReference+\DepthReference][c]{\Width}{\centering\strut#2}%
        }%
    }%
}

\begin{document}

\title{CoRT: Complementary Rankings from Transformers}

\author{Marco Wrzalik\qquad Dirk Krechel \\
  RheinMain University of Applied Sciences, Germany \\
  \texttt{\{firstname.lastname\}@hs-rm.de} \\}

\maketitle

\begin{tcolorbox}[colback=red!5!white,colframe=red!75!black]
\small
This work was published at NAACL-HLT 2021. For citations, please refer to the corresponding  conference proceedings:  \url{https://www.aclweb.org/anthology/2021.naacl-main.331/}
\end{tcolorbox}

\begin{abstract}

Many recent approaches towards neural information retrieval mitigate their computational costs by using a multi-stage ranking pipeline. In the first stage, a number of potentially relevant candidates are retrieved using an efficient retrieval model such as BM25. Although BM25 has proven decent performance as a first-stage ranker, it tends to miss relevant passages.
In this context we propose \projname{}, a simple neural first-stage ranking model that leverages contextual representations from pretrained language models such as BERT to complement term-based ranking functions while causing no significant delay at query time.
Using the MS MARCO dataset, we show that \projname{} significantly increases the candidate recall by complementing BM25 with missing candidates. Consequently, we find subsequent re-rankers achieve superior results with less candidates. We further demonstrate that passage retrieval using \projname{} can be realized with surprisingly low latencies.
\end{abstract}

\section{Introduction}
The successful development of neural ranking models over the past few years has rapidly advanced state-of-the-art performance in information retrieval \cite{guo19, craswell20}. 
One key aspect of the success is the exploitation of \emph{query-document interactions} based on token representations from self-supervised language models (LMs) \cite{devlin18, radford19, pennington14}. 
Due to high computational effort, however, these \emph{interaction-focused} approaches are limited to re-ranking scenarios and thus they depend on the effectiveness of first-stage ranking models for candidate retrieval. Term-based retrieval models such as BM25 have proven decent performance in this task, but tend to miss relevant passages.
In this context, we propose \emph{COmplementary Rankings from Transformers} (\projname{}), a simple neural first-stage ranking model that leverages contextual representations from \emph{transformer-based language models} \cite{vaswani17, devlin18} to complement term-based first-stage rankings. \projname{} optimizes an underlying text encoder towards representations that reflect relevance through vector similarity. The model is trained to act complementary to term-based retrieval by using passages from BM25 rankings as negative examples. 
We study the characteristics of \projname{} with four types of experiments based on the MS MARCO dataset. First, we measure various ranking metrics and compare the results with first-stage ranking baselines and competitors. In course of this, we demonstrate the portion of relevant candidates that are added by \projname{}. Second, we combine the candidates from \projname{} and BM25 with a state-of-the-art re-ranker based on BERT \cite{nogueira19} and investigate how many candidates are needed to saturate the ranking quality. Third, we train \projname{} with various representation sizes and measure its impact on the first-stage ranking quality. 
Fourth, we measure the retrieval latencies of \projname{} with two retrieval modalities: a distributed exhaustive search on four GPUs and an approximate search based on a graph-based nearest-neighbor index with pruning heuristics \cite{iwasaki18}. Finally, we build an exemplary end-to-end ranking pipeline using our first-stage ranking to demonstrate its efficiency. 
Our contribution is a first-stage ranking framework with the potential to improve end-to-end ranking pipelines by adding candidates that term-based retrieval models typically miss. With this it is possible to reduce the total number of re-ranking candidates without hurting end-to-end ranking quality. As a secondary contribution, we provide an open-source implementation\footnote{\url{https://github.com/lavis-nlp/CoRT}} that enables other researchers to reproduce our results and test \projname{} on other datasets.

\color{black}
\section{Background and Related Work}
In this section we describe key concepts of neural ranking and refer to related work. We then present neural first-stage ranking approaches that allow direct comparison with our results.

\subsection{Key Concepts of Neural Ranking} 
According to \citet{guo16}, Neural ranking approaches can be categorized into two types of models depending on the architecture. 
\emph{Representation-focused} approaches \cite{huang13, shen14, zamani18} produce representations for queries and documents to predict relevancy scores using a simple distance or similarity measure. 
In this context, exploiting \emph{local interactions} between neighboring terms is a commonly used technique \cite{shen14, zamani18}.
Models of the \emph{interaction-focused} type exploit interactions between query and document terms \cite{guo16, xiong17, dai18, nogueira19}. Although this leads to superior ranking quality \cite{guo16, guo19, qiao19}, it is computationally much more intensive, since a given query has to be processed together with each potentially relevant document. Hence, this type of neural ranking model is only applicable in a ranking pipeline, where limited numbers of documents are given as potentially relevant candidates. These candidates are selected by an efficient retrieval model that is able to retrieve documents directly from the corpus in a reasonable amount of time. The multi-stage ranking technique is also known as \emph{cascade ranking} and optimizing the configuration of such a pipeline towards maximized efficiency and effectiveness has been extensively studied in the past \cite{wang11, chen17}.
Many \emph{interaction-focused} neural ranking models employ a dedicated layer to explicitly perform a matching between query and document terms \cite{guo16, xiong17, dai18}. Another approach is using the \emph{attention mechanism} \cite{vaswani17}, or more specifically, a pretrained transformer encoder such as BERT \cite{devlin18} to exploit both local and query-document interactions \cite{nogueira19, qiao19}. 
Recently, some hybrid approaches have been proposed that combine typical representation-focused techniques with interaction-focused approaches to reduce computational cost: \citet{gao20} propose a model architecture comprising three modules for document understanding, query understanding and relevance judging respectively. The \emph{understanding modules} produce token-level representations, which can be cached as usual in representation-focused approaches. The \emph{relevance judging module} uses those cached representations to apply query-document interactions more quickly. Each module is a stack of transformer layers \cite{vaswani17}, initialized with weights from BERT. 
In a related approach, \citet{macavaney20} investigate the relationship between different numbers of dedicated layers of BERT for query-document interactions and measure the resulting speedup that is due to token representation caching, as well as its impact on the end-to-end ranking quality. 
\citet{khattab20} propose a related approach, namely \emph{ColBERT}. The model architecture incorporates an inexpensive max-similarity mechanism to perform token-level query-document interactions. The authors propose to store \emph{token} representations in an \emph{Approximate Nearest Neighbor} (ANN) index to quickly retrieve only those documents that have token representations in the proximity to those of the query. Thus, \emph{ColBERT} can be described as an end-to-end ranking approach that brings its own first-stage retrieval mechanism allowing to perform end-to-end ranking in a reasonable amount of time.

\subsection{Neural First-stage Ranking}
Now we discuss neural ranking approaches that can be used to retrieve passages or documents directly from an entire corpus in a reasonable amount of time and thus qualify for first-stage ranking. Many proposed methods make use of existing infrastructure for sparse bag-of-words retrieval or at least inverted indexing \cite{manning10}. \citet{zamani18} propose \emph{SNRM}, a representation-focused approach with sparse representations that can be used with an inverted index as if each feature dimensions corresponds to a term in a bag-of-words representation. SNRM uses pretrained \emph{GloVe Word Embeddings} \cite{pennington14} to model soft-matched n-grams which are encoded in sparse representations. \citet{nogueira19b} predict queries for given documents to expand those documents by corresponding query terms. In their first work, known as \emph{doc2query}, they used a sequence-to-sequence (seq2seq) transformer model \cite{vaswani17}. In a subsequent work, \citet{nogueira20} reported large effectiveness gains for their follow-up model \emph{docTTTTTquery} by replacing the seq2seq model with \emph{T5} \cite{raffel19}. Another approach aims at predicting optimal document term weights as a function of the term's context. \emph{DeepCT}, proposed by \citet{dai20}, utilizes BERT to predict these context-aware weights based on associated queries in the training data. 

Inverted indexing is only applicable to sparse representations. Representation-focused models using dense representations can instead employ an ANN index, which heuristically prunes documents that are unlikely to be in the top-$k$ proximity of the query representation to realize low response latencies \cite{boytsov16, gysel18}. \citet{karpukhin20} recently used this technique in combination with a fine-tuned BERT encoder for open question answering.

\section{Proposed Approach}
\newcommand{\simfunc}{sim}

We describe a first-stage ranking model that acts as a complementary ranker to existing term-based retrieval models such as BM25. To achieve this, we make use of a transformer-based pretrained language model and its inherent ability to make use of token-level local interactions. Its complementary behavior is further supported by negative sampling from BM25 rankings.

\subsection{Architecture}

The model architecture of \projname{}, illustrated in Figure \ref{fig:architecture}, follows the idea of a \emph{Siamese Neural Network} \cite{bromley93}. Passages and queries are encoded using an identical model with shared weights except for one detail: 
The passage encoder $\psi_{\alpha}$ and the query encoder $\psi_{\beta}$ use different segment embeddings \cite{devlin18}. \projname{} computes relevance scores as angular similarity between query and passage representations while training a \emph{pair-wise} ranking objective.

\subsection{Encoding} \label{seq:enc}

\projname{} can incorporate any BERT-like encoder as underlying text encoder. Here, we use a pretrained ALBERT \cite{lan19} encoder for its smaller model size, the tougher sentence coherence pretraining and increased first-stage ranking quality throughout our early-stage experiments compared to BERT.
The tokenizer of ALBERT is a WordPiece tokenizer \cite{wu16} including the special tokens \texttt{[CLS]} and \texttt{[SEP]} known from BERT.
From the text encoder we seek a single representation vector for the whole passage or query, which we call \emph{context representation}.
From ALBERT we take the \texttt{[CLS]} embedding of the last layer for this purpose. 
We denote the context representation obtained from the underlying encoder for an arbitrary string $s$ with $\tau(s) \in \mathbb{R}^h$ where $h$ is the output representation size. 

\begin{figure}[]
  \centering
  \includegraphics[width=\linewidth]{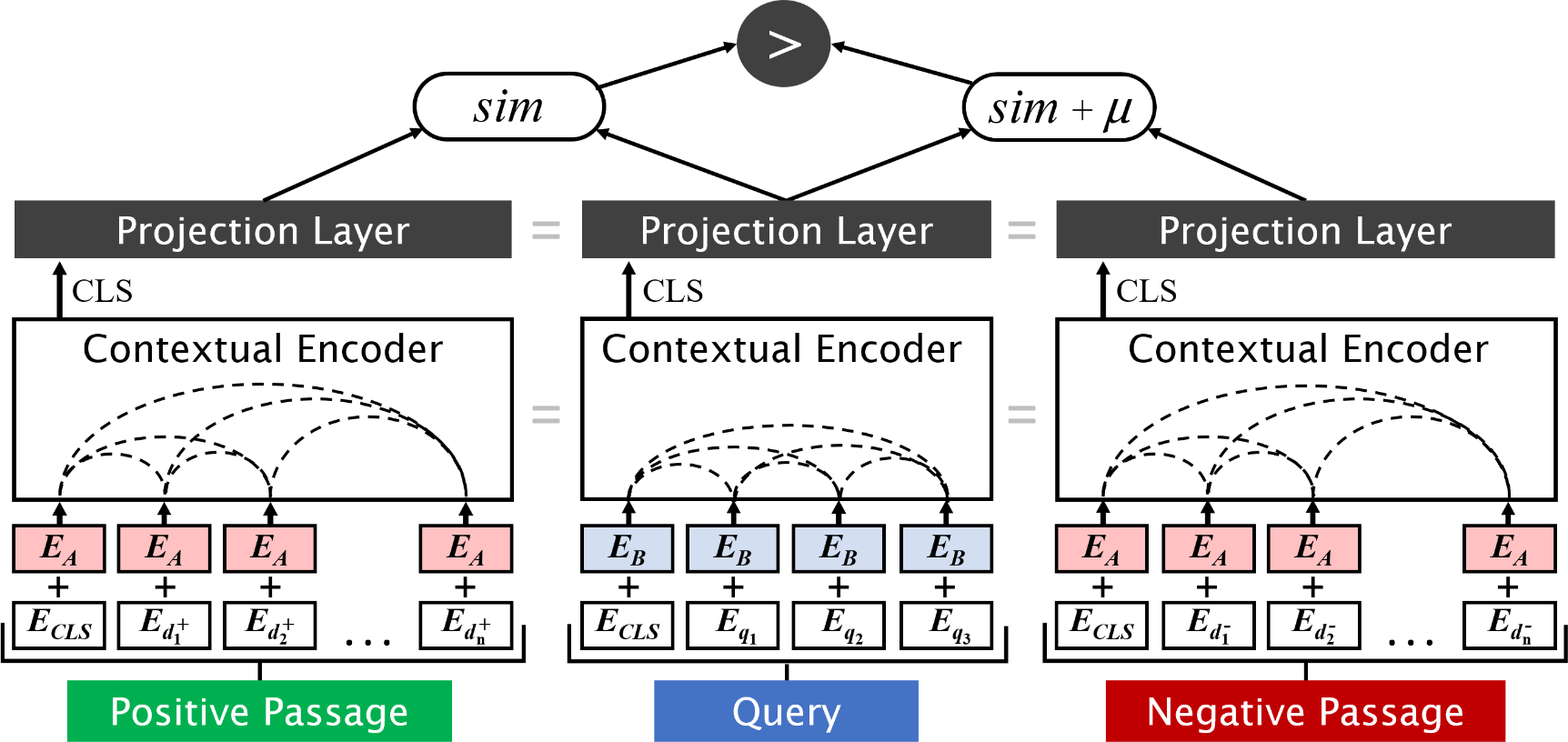}
  \caption{\projname{}'s model architecture and pair-wise learning objective (simplified).}
  \label{fig:architecture}
\end{figure}

ALBERT's language modeling approach involves sentence coherence prediction for which segment embeddings are used to signal different input segments. Although we only feed single segments to the encoder, i.e. a query or a passage, we use segment embeddings which allow the model to encode queries differently from passages.
The segment embeddings $E_A$ and $E_B$ (illustrated in Figure \ref{fig:architecture}) are part of the context encoder functions $\tau_{\alpha}$ and $\tau_{\beta}$ for passages and queries respectively. The context representation is further projected to the desired representation size $e$ using a linear layer followed by a $\tanh$ activation function.
Thus, the complete passage encoder function is $\psi_{\alpha}(s) := \tanh(W\tau_{\alpha}(s)+b)$ where $W\in \mathbb{R}^{h \times e}$ and $b \in \mathbb{R}^e$ are parameters of the linear layer. The query encoder $\psi_{\beta}$ is defined analogous.

\subsection{Training} \label{ssec:train}

Training \projname{} corresponds to updating the parameters of the encoder $\psi$ towards representations that reflect relevance between queries and passages through vector similarity. Each training sample is a triple comprising a query $q$, a positive passage $d^+$ and a negative passage $d^-$. While positive passages are taken from relevance assessments, negative passages are sampled from term-based rankings (i.e. BM25) to support the complementary property of \projname{}. The relevance score for a query-passage pair ($q, d$) is calculated using the angular cosine similarity function\footnote{Similar to \citet{cer18}, we found angular similarity performs better than cosine similarity.}:

\begin{small}
\begin{equation*}
\simfunc(q, d) := 1 - \arccos\left(\frac{ \psi_{\beta}(q)\cdot \psi_{\alpha}(d)}
                        {|| \psi_{\beta}(q)|| \; || \psi_{\alpha}(d)||}\right) / \pi
\end{equation*}
\end{small}

As illustrated in Figure \ref{fig:architecture}, the training objective is to score the positive example $d^+$ by at least the margin $\mu$ higher than the negative one $d^-$.
As part of our loss function, we use the triplet margin objective:

\begin{small}
\begin{equation*} \label{eq:l}
l(q, d^+, d^-) := max(0, \simfunc(q, d^-) - \simfunc(q, d^+) + \mu)
\end{equation*}
\end{small}

Inspired by \citet{oh16}, we aim to take full advantage of the whole training batch.
For each query, each passage in the batch is used as a negative example, except for the given positive passage. Thus, we define our batch-wise loss function as follows:

\begin{small}
\begin{equation*}
    \mathcal{L} :=
    \sum\limits_{1\leq i \leq n} 
    \left(\sum\limits_{1\leq j \leq n}
    l(q_i, d^+_i, d^-_j) \; + \hspace{-0.3cm} 
    \sum\limits_{1\leq k \leq n,\; k \neq i}
    \hspace{-0.3cm} l(q_i, d^+_i, d^+_k)\right)
\end{equation*}
\end{small}

$q_i$, $d^+_i$ and $d^-_i$ denote the triple of the $i_{th}$ sample in the batch and $n$ the number of samples per batch. We found this technique to make the training process more robust against exploding gradients. Otherwise we need to employ gradient clipping \cite{zhang19} to stabilize the training process. Also, it positively affects first-stage ranking results\footnote{We achieve 2.0 p.p. higher MRR@10 compared to the plain triplet margin loss on the MS MARCO passage task.}.

\subsection{Indexing and Retrieval} \label{ssec:retrieval}
For retrieval with \projname{}, each passage must be encoded by the passage encoder $\psi_{\alpha}$. Subsequent normalization of each vector allows us to use the dot product as a proxy score function for $\simfunc{}$, which is sufficient to form rankings accurately. Given a query $q$, we calculate its representation $\psi_{\beta}(q)$ and the dot product with each normalized passage vector. From those, the $k$ highest scores are selected and sorted to form the \projname{} ranking. This procedure can be implemented heavily parallelized using GPU matrix operations. Alternatively, the passage representations can be indexed in an ANN index to avoid exhaustive similarity search. In contrast to the first-stage ranking of \citet{khattab20} and \citet{macavaney20}, we only index one representation per passage rather than one per token.
Finally, we combine the resulting ranking of \projname{} with the respective BM25 ranking by \textit{interleaving} the positions beginning with \projname{} to create a single merged ranking of equal length. During this process, each passage that was already added by the other ranking is omitted.
For example, merging two ranking lists beginning with $[a,b,c,d,\dots]$ and $[e,c,f,a,\dots]$ would result in $[a,e,b,c,\cancel{c},f,d,\cancel{a},\dots]$. The interleaving procedure stops as soon as the desired ranking size has been reached. The result is a compound ranking of \projname{} and BM25, which we denote with \projname{BM25}.

\section{Experiments}

\begin{table*}
  \centering
    \caption{First-stage ranking results on the MS MARCO \texttt{dev.small} set. The asterisk (*) denotes merged rankings using an instance of CoRT that was not specifically trained to complement the corresponding term-based ranker.}
  \label{tab:results}
  \small
  \begin{tabu}{X[2.5,l]|X[c]|X[c]|X[c]X[c]X[c]X[c]X[c]}
    \toprule

    MS MARCO  & MRR & NDCG & \multicolumn{5}{c}{RECALL} \\
    Passage (\texttt{dev.small})
             & \multicolumn{1}{c|}{@10}
             & \multicolumn{1}{c|}{@20}
             & \multicolumn{1}{c}{@50}
             & \multicolumn{1}{c}{@100}
             & \multicolumn{1}{c}{@200}
             & \multicolumn{1}{c}{@500}
             & \multicolumn{1}{c}{@1000} \\
    \midrule
    BM25                              & 18.7 & 25.8 & 59.2 & 67.0 & 73.8 & 81.2 & 85.7 \\
    doc2query      & 21.5 & - & - & - & - & - &  89.3 \\
    DeepCT  & 24.3  & 32.1 & 68.5 & 75.2 & 81.0 & 87.3 & 90.9  \\
    docTTTTTquery  & 27.7 & 36.5 & 75.6 & 81.9 & 86.9 & 91.6 & 94.7 \\
    \midrule
    \projname{}                       & 27.1 & 34.0 & 66.4 & 73.1 & 78.6 & 84.4 & 88.0 \\
    \projname{}\textsubscript{BM25}   & 27.4 & 35.9 & 74.3 & 81.6 & 87.3 & 92.5 & 94.9  \\
    \projname{DeepCT*}               & 28.3 & 36.8 & 75.6 & 82.3 & 87.6 & 92.5 & 94.9  \\ 
    \projname{DocTTTTTquery*}         & \textbf{28.8} & \textbf{38.0} & \textbf{78.5} & \textbf{85.3} & \textbf{90.0} & \textbf{94.4} & \textbf{96.5} \\ 
    \bottomrule
  \end{tabu}
\end{table*}
\normalfont

We present four experiments studying the ranking quality and recall of \projname{}, the connection between the number of candidates and re-ranking effectiveness, the impact of the representation size $e$, and \projname{}'s retrieval latencies. Finally, we outline a competitive end-to-end ranking setup with \projname{} and a BERT-based re-ranker.

\subsection{Datasets} \label{s:datasets}
\mysubsub{MS MARCO Passage Retrieval}
The \emph{Microsoft Machine Reading Comprehension} \cite{bajaj16} dataset for passage ranking was introduced in 2018. It provides a benchmark for passage retrieval with real-world queries and passages gathered from Microsoft's \emph{Bing} search. The MS MARCO passage ranking task comprises 8.8M passages sampled from web pages and about 1M queries that are formulated as questions. The objective is to rank those passages high that were labeled as relevant to answer the respective question. The annotations, however, are sparse. There are 530k positive relevance labels distributed over 808k queries in the \texttt{training} set, whereby most queries are associated to one passage.
The validation and evaluation sets, \texttt{dev} and \texttt{eval}, comprise 101k queries each. An official subset of \texttt{dev}, called \texttt{dev.small} comprises 6980 queries and 7437 relevance labels and is often used for publicly reported evaluations. We follow this convention and use \texttt{dev.small} for testing.
The creators suggest to use the \emph{mean reciprocal rank} cut at the tenth position (MRR@10) as primary evaluation measure. Additionally, we measure NDCG@20 \cite{manning10} as less punishing ranking quality measure and the recall at various positions to indicate how many relevant passages a re-ranker would miss if the number of candidates is reduced.

\mysubsub{TREC 2019 DL Passage Retrieval}
The passage retrieval section of the \emph{TREC 2019 Deep Learning Track} \cite{craswell20} provides on average 215 manual relevance assessments per query for a set of 43 MS MARCO queries. Each assessment corresponds to a rating on a scale from 0 (not relevant) to 3 (perfectly relevant). We adopt the evaluation metrics MRR (uncut), NDCG@10 and MAP from the official TREC overview. In contrast to the original MS MARCO benchmark, this evaluation set provides dense annotations, but only for few queries.

\subsection{First-Stage Ranking}

We train \projname{} as described in Section \ref{ssec:train} while using a representation size of $e=768$.
In this section we discuss the first-stage ranking results of our model using the datasets and their associated metrics described in Section \ref{s:datasets}.

\mysubsub{MS MARCO Passage Retrieval}
The results of \projname{} and its baselines on the MS MARCO passage retrieval task (\texttt{dev.small}) are reported in Table \ref{tab:results}. Next to BM25 as a baseline, we include \emph{DeepCT} \cite{dai20},  \emph{doc2query} \cite{nogueira19b} and its successor \emph{docTTTTTquery} \cite{nogueira20}.
All three are recent first-stage rankers with average retrieval latencies below 100ms per query on the MS MARCO passage corpus. The metrics MRR@10 and NDCG@20 reveal a quite decent ranking quality for the standalone \projname{} ranker. Since \projname{}'s primary use is candidate retrieval rather than standalone ranking, we pay particular attention to the recall at various cuts. From the perspective of BM25, the absolute increase of recall due to merging with \projname{} ranges between 15.1 (RECALL@50) and 9.2 (RECALL@1000), which we consider the complementary portion of \projname{}. Greater increases of recall can be noticed for lower cuts, which is particular useful when re-ranking is performed with low numbers of candidates. The top-200 candidates from \projname{BM25} comprise higher recall than the top-1000 candidates from BM25. 
The metrics for \emph{DeepCT} and \emph{docTTTTTquery} have been calculated using published top-1000 rankings from the respective authors. Thus, we were able to merge those rankings with \projname{}. However, the used instance of \projname{} was trained on BM25 and not on the external ranker.

\begin{table}
  \caption{First-stage ranking results on the \emph{TREC 2019 DL} passage task.}
  \label{tab:trec19}
  \centering
  \small
  \begin{tabu}{X[1.4,l]X[c]X[c]X[c]X[c]}
    \toprule
    TREC DL & MRR & NDCG & MAP & RECALL \\
    2019
             & \multicolumn{1}{c}{@1000}
             & \multicolumn{1}{c}{@10}
             & \multicolumn{1}{c}{@1000}
             & \multicolumn{1}{c}{@500} \\
    \midrule

    BM25               & 68.5 & 49.7 & 29.0 & 69.4 \\
    \projname{}        & 84.3 & \textbf{60.0} & 29.7 & 58.3 \\
    \projname{BM25}    & \textbf{86.2} & 59.7 & \textbf{35.1} & \textbf{76.9} \\
    
    \bottomrule
  \end{tabu}
\end{table}

\mysubsub{TREC 2019 DL Passage Retrieval}
Although we consider the relevance assessments from TREC 2019 DL to be dense, we found 112 unlabeled passages among the 43 top-10 \projname{} rankings while the assessments for the BM25 rankings are complete. This means there are, on average, 2.6 unlabeled passages in the top-10 \projname{} rankings, which might make this evaluation somewhat unfavourable for \projname{} and explain the drop in recall. Still, Table \ref{tab:trec19} shows superior results for \projname{} compared to BM25 in terms of ranking quality (MRR, NDCG and MAP). Merging \projname{} with BM25 slightly increases MRR and NDCG, while a decent gain in terms of recall can be noticed. We can not report any results for \emph{DeepCT} or \emph{docTTTTTquery}, since only rankings for the MS MARCO \texttt{dev.small} set are available online from the respective authors.

\subsection{Candidate Re-ranking} \label{s:reranking}

We re-rank candidates from both BM25 and \projname{BM25} to study the impact of the candidates on a subsequent \emph{interaction-focused} re-ranking. By varying the numbers of candidates, we investigate at which point adding more candidates becomes ineffective.

\mysubsub{Re-ranking Model} Similar to \cite{nogueira19}, we use a simple binary classifier based on BERT. 
The query-passage pair $(q,p)$ is concatenated to one token sequence of two segments. This sequence is processed by the BERT encoder while the \texttt{[CLS]} embedding of the last layer, which we denote with $\phi(q,p)$, is projected to a single classification logit. We then apply the sigmoid activation function $\sigma$ to obtain the relevance confidence for query $q$ and passage $p$. This procedure can be formalized as $\zeta(q,p)=\sigma(W'\phi(q,p)+b')$ where $W'\in \mathbb{R}^{h \times 1}$ and $b' \in \mathbb{R}$ are the parameters of a linear layer with a single output activation. To form a ranking at inference time, we sort the candidates by the model's confidence.   
Following \cite{nogueira19}, this model is trained using a point-wise objective. We sample query-passage pairs, each associated with a binary relevance label $y\in\{0,1\}$ and minimize the binary cross-entropy loss:
\begin{equation*}
l'(q,p,y) = y \cdot log \; \zeta(q,p)   + (1-y) \cdot log \; (1-\zeta(q,p))
\end{equation*}

\begin{figure}
  \begin{center}
    \resizebox{\linewidth}{!}{\clip{
      \input{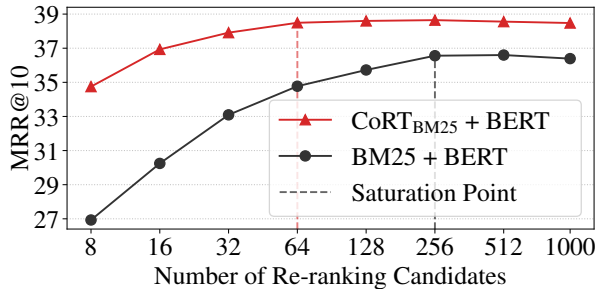}
    }}
  \end{center}
  \caption{Re-ranking quality by number of candidates for BM25 (black) and \projname{BM25} (red) on the MS MARCO passage task. Dashed lines indicate effectiveness saturation ($<$0.5\% increase).}
  \label{fig:topx}
\end{figure}

\mysubsub{Re-ranking Results}
As illustrated in figure \ref{fig:topx}, using \projname{BM25} as first-stage ranking appears to result in superior end-to-end ranking quality in terms of MRR@10. This is especially true, if low numbers of candidates are used.
We also notice earlier saturation\footnote{In our definition, saturation is reached, when doubling the number of candidates results in less than $0.5\%$ increase of the respective metric} of MRR@10 for \projname{BM25}, which is illustrated in Figure \ref{fig:topx}. Only 64 candidates from \projname{BM25} are sufficient to achieve top results with this re-ranker. In contrast, 256 candidates from BM25 are needed to reach the point of saturation, which translates in quadrupled re-ranking time.

\begin{table}[b]
  \caption{First-stage ranking results for various representation sizes.}
  \label{tab:esize}
  \centering
  \small
  \begin{tabular}{l|c|c|c|cc}
    \toprule
    && MRR & nDCG &  \multicolumn{2}{c}{RECALL} \\
     & $e$
        & \multicolumn{1}{c|}{@10}
        & \multicolumn{1}{c|}{@20}
        & \multicolumn{1}{c}{@200}
        & \multicolumn{1}{c}{@1k}\\
    \midrule
    \multirow{4}{*}{\projname{}}
    & 32 & 23.6 & 29.8 & 70.6 & 81.8\\
    & 64 & 25.6 & 32.3 & 75.6 & 85.7 \\
    & 128 & 26.8 & 33.4 & 77.2 & 87.1  \\
    & 256 & 26.8 & 33.6 & 78.2 & 87.8 \\
    & 768 & 27.1 & 34.0 &  78.6 & 88.0 \\
    \midrule
    \multirow{4}{*}{\projname{BM25}} 
    & 32 &  25.2 & 33.5 & 85.2 & 93.7\\
    & 64 &  26.6 & 34.9 & 86.2 & 94.4\\
    & 128 &  27.4 & 35.7 & 87.0 & 94.8 \\
    & 256 &  27.2 & 35.6 & 87.0 & 94.9 \\
    & 768 &  27.4 & 36.0 & 87.3 & 94.9 \\
    \bottomrule
  \end{tabular}
\end{table}

\subsection{Impact of Representation Size} \label{ssec:repr-size}

As described in Section \ref{seq:enc}, \projname{} projects the context representation of the underlying encoder $\tau$ to an arbitrary representation size $e$. This size determines the size of the final index and also influences the retrieval latency. 
The total size of the encoded corpus is easy to calculate. For example, with $e=128$ and the MS MARCO corpus, the index size (without overhead) amounts $8.8M\;documents \times 128 floats/document \times 4 bytes/float \approx 4.5 \times 10^9 bytes \approx 4.2GB$. Thus, $e$ is proportional to the total size and reducing $e$ to 64 would halve the memory footprint.
If $e$ is small, however, it is more difficult to attain the training objective. Thus, $e$ can be used for a trade-off between ranking quality and computational effort / resource cost. We investigate the relation between the representation size $e$ and the ranking quality by conducting identical training runs with different numbers for $e$. The results in Table \ref{tab:esize} show that MRR@10 already saturates at $e=128$. Interestingly, even with an representation size of $e=32$, \projname{} outperforms BM25 in terms of MRR@10 and nDCG@20 by a big margin.

\subsection{Latency Measurements}

We propose two methods for the deployment of \projname{}. The first exhaustively calculates similarity scores using multiple GPUs while the second incorporates an \emph{Approximate Nearest Neighbor} index (ANN). We measure retrieval latencies of those methods and compare them with BM25 as representative for term-based retrieval models based on inverted indexing. We conduct the latency measurement based on the top-1000 retrieval for the \texttt{dev.small} split of the MS MARCO passage corpus. Since some approaches profit from batch computing, we also measure the latency for batches of 32 queries. As representation size, we have chosen $e=128$, since it is the smallest representation size investigated in Section \ref{ssec:repr-size} that does not hurt the ranking quality of \projname{BM25}.

 \mysubsub{Lucene BM25 Baseline}
As retrieval latency baseline, we use a \emph{Lucene} index generated by the \emph{Anserini} toolkit \cite{yang17}. 
Please note, this is not perfectly representative for sparse bag-of-words retrieval: Retrieval latency can be reduced due to index pruning without significantly hurting retrieval quality  \cite{mackenzie20}. The retrieval was performed on a machine with an Intel Core i9-9900KS processor (16 logical cores, 8 physical) and enough memory to hold the whole corpus. Single queries were processed using the single-threaded search function, while batch-wise search has been performed with 16 threads. 

\mysubsub{Retrieval using multiple GPUs}
Multiple GPUs can be used to deploy \projname{} for fast large-scale ranking. 
We propose to uniformly distribute the vector representations of the corpus on the available GPUs. Each GPU ranks its own partition of the corpus as described in Section \ref{ssec:retrieval}. Afterwards, the results for each partition are aggregated by selecting the top-k candidates with highest scores.

\mysubsub{Retrieval using ANN}
Since \projname{} operates on vector similarities, it can make use of ANN search. We measure the retrieval latency and the loss of recall, which occurs due to the pruning heuristics. We use a graph-based index optimized with the \emph{ONNG} method \cite{iwasaki18}. An implementation of this method is publicly available as part of the \emph{NGT Library}\footnote{\url{https://github.com/yahoojapan/NGT}}. To adjust the trade-off between retrieval latency and accuracy of the index, we alter the search range coefficient $\epsilon$. We always retrieve 1000 candidates from the ANN index, even if we use a smaller number of candidates in a ranking pipeline. 

\begin{table}
  \caption{Retrieval latencies averaged over 6980 queries of the \emph{dev.small} split.}
  \label{tab:latency}
  \centering
  \small
  \begin{tabu}{X[3.0,l]|X[c]|X[c]X[c]}
    \toprule
    MS MARCO  & RECALL & \multicolumn{2}{c}{LATENCY (ms)} \\
    Passage (\texttt{dev.small})
        & \multicolumn{1}{c|}{@200}
        & \multicolumn{1}{c}{Single}
        & \multicolumn{1}{c}{Batch32}\\
    \midrule
    BM25 (anserini) &  73.8 & 38 & 290 \\
    \midrule
    \textbf{\projname{}} ($e=128$) & & &  \\
    \; \underline{Query Encoding} &&& \\
    \; \; - Single GPU & - &  8 & 17 \\
    \; \underline{Retrieval} &&& \\
    \; \; - Single GPU &  77.2 & 68 & 164 \\
    \; \; - Quad GPU &  77.2 & 17 & 35 \\
    \; \; - ANN\textsubscript{$\epsilon = 0.01$}  & 76.6 & 4  & - \\
    \; \; - ANN\textsubscript{$\epsilon = 0.1$} & 76.9 & 17 & -  \\
    \; \; - ANN\textsubscript{$\epsilon = 0.4$}  & 77.2 & 71  & - \\
    \midrule
    \textbf{\projname{BM25} Total} &&& \\
    \; - Quad-GPU & 87.0 &  $\sim$63 & $\sim$342  \\
    \; - ANN\textsubscript{$\epsilon = 0.1$} & 86.9 &  $\sim$63 & -  \\
    \; - ANN\textsubscript{$\epsilon = 0.01$} & 86.8 &  $\sim$50 & -  \\
    \bottomrule
  \end{tabu}
\end{table}

\mysubsub{Latency Measurements} The latency measurements are reported in Table \ref{tab:latency}. For \projname{} the total retrieval latency per query consists of two factors: Query encoding and retrieval. The query encoding has to be performed by the query encoder $\psi_{\beta}$, which we highly recommend to run on a GPU. The latency of the retrieval depends on the retrieval methods described above. Employing multiple GPUs appears to reduce retrieval latency on a linear scale: The exhaustive search using four GPUs takes 17ms for a single query, while a single GPU takes four times as long. The total retrieval time per query sums up to $17+8=25ms$ for the quad GPU setting, which is below the BM25 baseline.
It is worth noting, that batch-wise processing results in a substantial efficiency increase: 
Retrieval for 32 queries at once only takes about twice as long as a single query. This can be useful if multiple queries queue up while the system is busy.
The tested BM25 index (Anserini), on the other side, seems to suffer from multiprocessing overhead or some sort of bottleneck.
The latencies for the ANN index has been measured with three different values for the search range coefficient $\epsilon$. While this significantly affects the retrieval latency, only slight differences regarding the quality of the first-stage ranking are observed. Latencies for \projname{BM25} comprise latencies from  BM25, \projname{}’s  query  encoding and the corresponding retrieval method. For simplicity, we assume sequential processing of all three components, although BM25 could be processed in parallel. 


\subsection{End-to-end Retrieval}

\begin{table}[b]
  \caption{Measured end-to-end ranking quality and latency. * ColBERT's latency was measured with different hardware}
  \label{tab:end-to-end}
  \centering
  \small
  \begin{tabu}{X[2.6,l]X[0.8,c]X[c]}
    \toprule
    MS MARCO  & MRR  & \multicolumn{1}{c}{LATENCY} \\
    Passage (\texttt{dev.small})
        & \multicolumn{1}{c}{@10}
        & \multicolumn{1}{c}{(ms)}\\
    \midrule
    ColBERT\textsubscript{L2} \cite{khattab20} & 36.0   & \;458* \\
    \midrule
    \projname{}\textsubscript{BM25} (ANN\textsubscript{$\epsilon = 0.1$}, top-64) + BERT Re-ranking & 38.4  &  255 \\
    \bottomrule
  \end{tabu}
\end{table}

Intrigued by the remarkable ratio of retrieval latency and ranking quality of ColBERT's full-ranking approach \cite{khattab20}, we used our above findings to create a competitive end-to-end ranking setup. We suggest to re-rank the top-64 candidates from \projname{BM25} with $e=128$, retrieved by an ANN index ($\epsilon = 0.1$). The end-to-end latency comprises $8ms$ for query encoding, $17ms$ for \projname{} retrieval based on ONNG, $38ms$ for BM25 retrieval, and $192ms$ for re-ranking. Although the BM25 candidates could be retrieved in parallel, we report sequential processing latencies. As shown in Table \ref{tab:end-to-end}, we outperform ColBERT's end-to-end ranking performance in terms of MRR@10 and retrieval latency. It is worth noting that the "RTX 2080 Ti" we used for latency measurements is less powerful than the "Tesla V100" \citet{khattab20} used for their measurements. \projname{}'s representations for the MS MARCO corpus only weight 4.3GB when $e$ is set to 128, or 7.0GB when indexed in an ONNG index. The size of the query encoder only amounts about 50MB, which is due to ALBERT's parameter sharing. To compile the full \projname{BM25} candidates, the corresponding BM25 index is needed, which amounts 2.2 GB on disk. Although more memory is needed to deploy and operate both indexes, this is by far less than the 154GB footprint reported by \citet{khattab20} for ColBERT's end-to-end approach.

\section{Conclusion}
In this paper, we propose \projname{}, a framework and neural first-stage ranking model that leverages contextual representations from transformer-based language models to complement term-based ranking functions. As a result, we observe decently increased recall measures and improved end-to-end ranking quality on the MS MARCO passage task. Also, we are able to decrease the number of candidates for re-ranking without hurting the final performance. Our further experiments reveal sweet spots for \projname{}'s representation size and the number of re-ranking candidates. We presented two deployment strategies for \projname{} and measured their performances in terms of efficiency and effectiveness. Finally, we demonstrate \projname{} can be used with a simple BERT-based re-ranker to create a competitive ranking pipeline.

\section*{Acknowledgments}
We would like to thank Felix Hamann and Prof. Dr. Adrian Ulges for helpful discussions and comments on the manuscript, as well as the anonymous reviewers for their valuable feedback. This work was funded by German Federal Ministry of Education and Research (Program FHprofUnt, Project DeepCA (13FH011PX6)).
 
\balance

\bibliographystyle{acl_natbib}
\bibliography{paper}




\appendix

\onecolumn
\begin{multicols}{2}
\section{Appendix: Implementation Details} \label{s:impl}
\subsection{Hardware \& Software} We use up to four NVIDIA GTX 2080 TI graphic cards in combination with 128GB DDR4 RAM and an Intel Core i9-9900KS processor. We use \emph{PyTorch} \cite{pytorch19} and \emph{HuggingFace's Transformers} \cite{wolf19} as deep learning libraries. BM25 rankings are generated using the \emph{Anserini} toolkit \cite{yang17}.

\subsection{\projname{} Training}
We train \projname{} based on the pretrained ALBERT model "\texttt{albert-base-v2}", which is the lightest available version in \emph{HuggingFace's} repository\footnote{\url{https://huggingface.co/transformers/pretrained_models.html}}. Each model is trained for 10 epochs, where each epoch includes all queries that are associated to at least one relevant document plus one randomly sampled positive and one negative passage. 
Negative examples are sampled from unlabeled passages of top-100 BM25 rankings. There, we exclude the first 8 ranks to reduce the probability of drawing actual relevant passages and thus give contradictory signals less often. 
We find this slightly increases \projname{}'s ranking quality.
As usual for BERT-based models we use the ADAM optimizer with weight decay fix \cite{adamw17} and the default parameters $\beta_1 = 0.9$, $\beta_2=0.999$, $eps=10^{-6}$, a weight decay rate of $\lambda=0.1$ and a linearly decreasing learning rate schedule starting with $lr=2\times10^{-5}$ after 2.000 warm-up steps. We train mini-batches of size $n=6$ samples (triples) while accumulating the gradients of 100 mini-batches before performing one update step. The triplet margin is set to $\mu=0.1$, which has been tuned within the range of $[0.01, 0.2]$.

\subsection{Re-ranker Training}
Our BERT re-ranking experiment utilizes the pretrained "\texttt{bert-base-uncased}" model, hosted by \emph{HuggingFace}. We use equal optimizer settings than for \projname{} except for the learning rate, which we empirically set to $5\times10^{-5}$. We use a batch-size of 8 and accumulate the gradients of 16 batches.

\section{Appendix: Retrieval Examples}

Table \ref{tab:examples} shows top-1 retrieval examples of \projname{} and BM25. The first query exemplifies the advantage of local interactions in the query encoder. We hypothesize, the query could be interpreted as a question about the density of aluminum although the term density was not included. The second query is an example, where BM25 works well due to favorable keywords in the passage. Although \projname{}'s top result is not labeled, it clearly is relevant to the query. Since the passage misses the keyword "insane", it is difficult to retrieve for a term-based model. We hypothesize, due to the terms "hallucinations" and "paranoia", \projname{} is able to match the contexts in this example.

\end{multicols}

\begin{table}[!htb] 
  \caption{Retrieval Examples with highlighted keywords. Ranks beyond the top-1000 are denoted with "n/a".}
  
  \label{tab:examples}
  \centering
  \newcolumntype{C}{>{\centering\arraybackslash}p{2em}}
  \small

  \begin{tabular}{|m{4.5em}|m{33em}|C|C|C|}
    \toprule
    \bf\multirow{2}{*}{Query} &
    \bf\multirow{2}{*}{Sample Passage} & 
    \bf\multirow{2}{*}{Label} & 
    \multicolumn{2}{c|}{\bf Rank}   \\
    & & &  BM25 & \projname{} \\
    
    \midrule
        
        \multirow{4}{*}{\parbox{4.5em}{how much does \blue{aluminum} \pink{weigh}}} &
        Question and answer. how much does a western 14 ft.\blue{aluminum} boat \pink{weigh}? what is the weight difference between a 12 ft. and 14 ft. \blue{aluminum} boat. 12 ft \blue{aluminum} boat with no gear or anything would probably \pink{weigh} about 115-150 lbs, 14 ft \blue{aluminum} boat with no gear or anything would probably \pink{weigh} about 250-300 lbs. \vspace{0.1em} &
        n/a &    1 &    n/a \\

        \cline{2-5}

        & \vspace{0.1em} Quick Answer. One cubic inch of \blue{aluminum} \pink{weighs} 1.56 ounces. The metal sinks in water, but it is still relatively lightweight. The density of \blue{aluminum} is 2.7 grams per milliliter. \blue{Aluminum} is used as a metal foil, a conductor of electricity and in the construction of airplane fuselages. & True & 735 & 1 \\
    \midrule

        \multirow{4}{*}{\parbox{4.5em}{how many \green{days} of no \blue{sleep} until \pink{insane}}} &
        Although there are many articles stating one can be declared legally \pink{insane} if... How long must you go without \blue{sleep} to be declared legally \pink{insane} Each state would have different laws regarding the requirements of declaring... Is it true that after three \green{days} of complete \blue{sleep} deprivation you are considered legally \pink{insane}? Not after 3 \green{days} but any longer will. \vspace{0.1em} &
        True &    1 &    n/a \\

        \cline{2-5}

        & \vspace{0.1em} The longest recorded time a human has ever gone without \blue{sleep} is 18 \green{days}, 21 hours, and 40 minutes, which resulted in hallucinations, paranoia, etc. However most people can only last 4-6 \green{days} without stimulants, and about 7-10 \green{days} before the body will be unable to function and long term damage can be caused. & n/a & n/a & 1 \\
   
    \midrule
  \end{tabular}
  \vspace{-2em}
\end{table}




\end{document}